\documentclass[11pt,twoside]{article}
\usepackage{asp2010}

\resetcounters

\begin{document}
\title{MHD simulations reveal crucial differences between solar and very-cool star magnetic structures}
\author{Benjamin~Beeck$^{1,2}$, Manfred~Sch\"ussler$^1$, and Ansgar~Reiners$^2$}
\affil{$^1$Max Planck Institute for Solar-System Research, Max-Planck-Stra\ss e 2, 37191 Katlenburg-Lindau, Germany}
\affil{$^2$Georg August University, Institute for Astrophysics, Friedrich-Hund-Platz 1, 37077 G\"ottingen, Germany}

\begin{abstract}
We carried out 3D radiative magnetohydrodynamic simulations of the convective and magnetic structure in the surface layers (uppermost part of the convection zone and photosphere) of main-sequence stars of spectral types F3 to M2. The simulation results were analyzed in terms of sizes and properties of the convection cells (granules) and magnetic flux concentrations as well as velocity, pressure, density, and temperature profiles. 
\\Our numerical simulations show for the first time a qualitative difference in the mag\-ne\-to-convection between solar-like stars and M dwarfs. Owing to higher surface gravity, lower opacity (resulting in higher density at optical depth unity), and more stable downflows, small-scale magnetic structures concentrate into pore-like configurations of reduced intensity. This implies that in very cool stars magnetic surface structures like plage regions and starspots significantly differ from the solar example. Such a difference would have major impact on the interpretation of Doppler imaging data and the analysis of M dwarf spectra.
\end{abstract}

\section{Introduction}
Cool main-sequence stars of spectral types F through L have a thick convective envelope or are fully convective. In many of such stars, magnetic fields of various strengths have been detected.  In the Sun, the surface magnetic field is observed to be highly structured owing to its interaction with the convective flows. This has significant impact on the magnetic signatures in spectral lines that are used to detect and measure the field. In contrast to the Sun, the structure and properties of magnetic fields on other cool stars are unknown. In the absence of spatially resolved observations, the effect of the magnetic structure on signatures of the magnetic field can be evaluated by numerical simulations of the magneto-convective processes.

\section{The MURaM code and simulation setup}

\begin{figure}
  \begin{center}
    \begin{tabular}{cc}
      {\bf\large F3V} & {\bf\large G2V} \\
      \includegraphics[width=0.40\linewidth, trim=0mm -1mm 0mm 0mm]{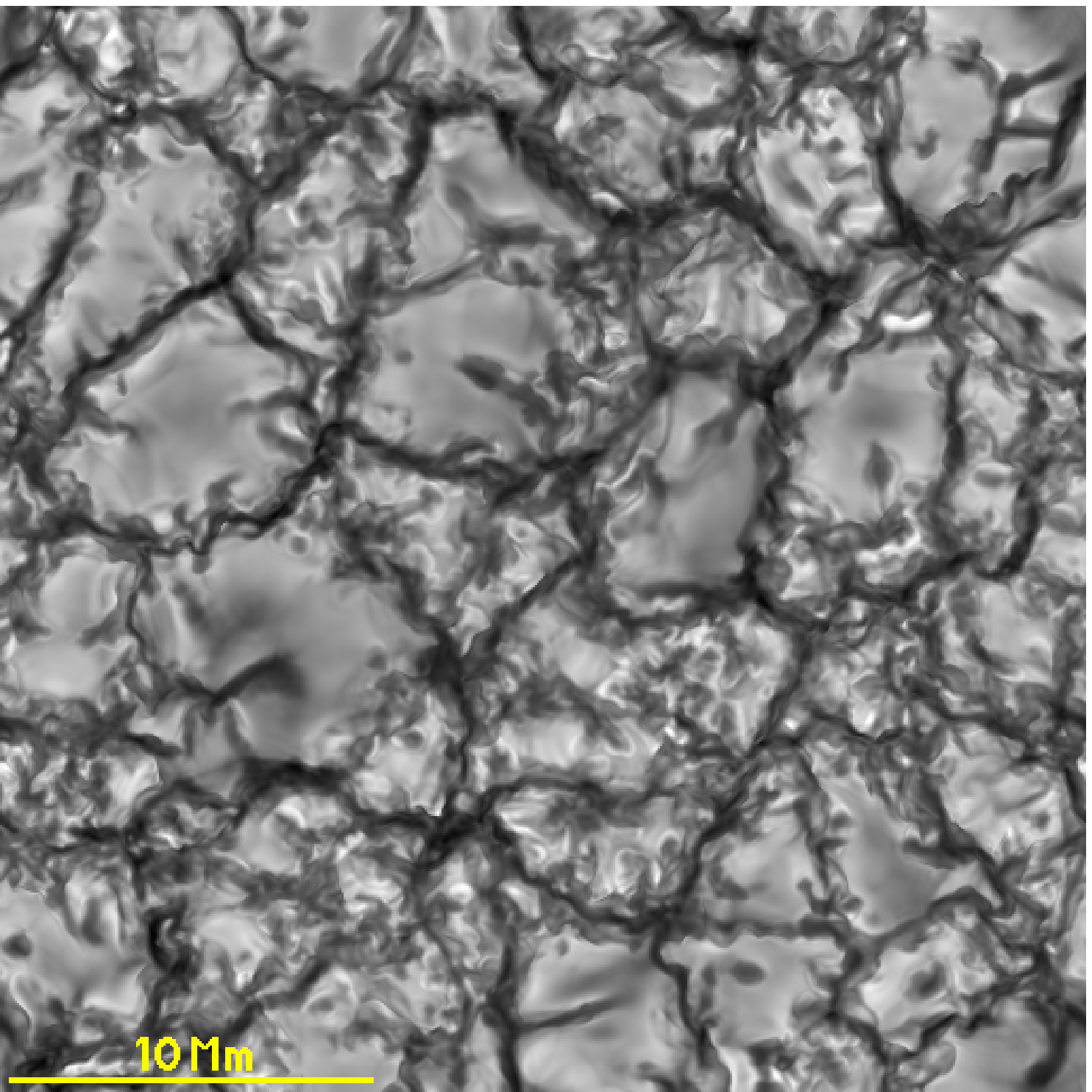} & \includegraphics[width=0.40\linewidth, trim=0mm -1mm 0mm 0mm]{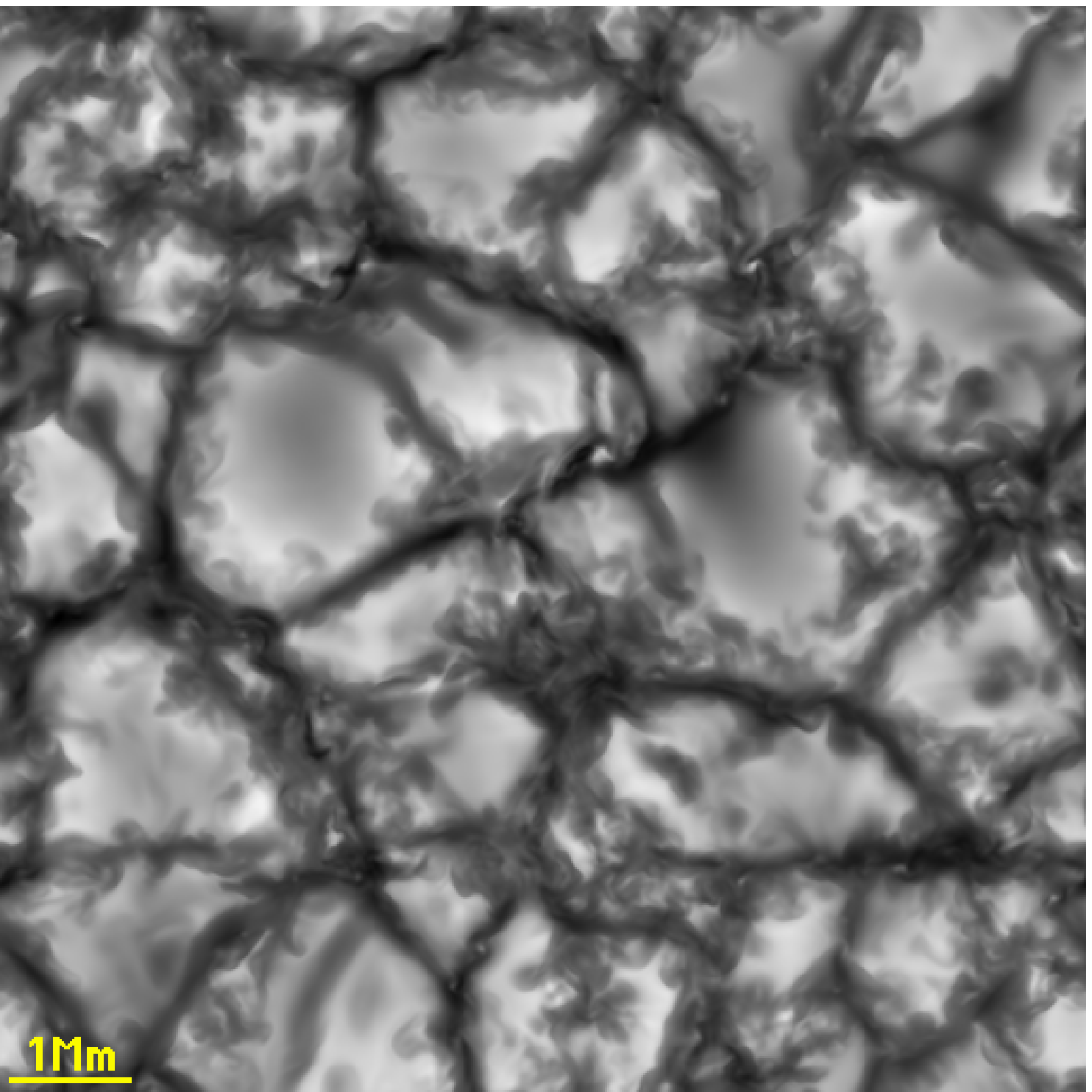}\\\hline
                      {\bf\large K0V}  & {\bf\large K5V} \\
                      \includegraphics[width=0.40\linewidth, trim=0mm -1mm 0mm 0mm]{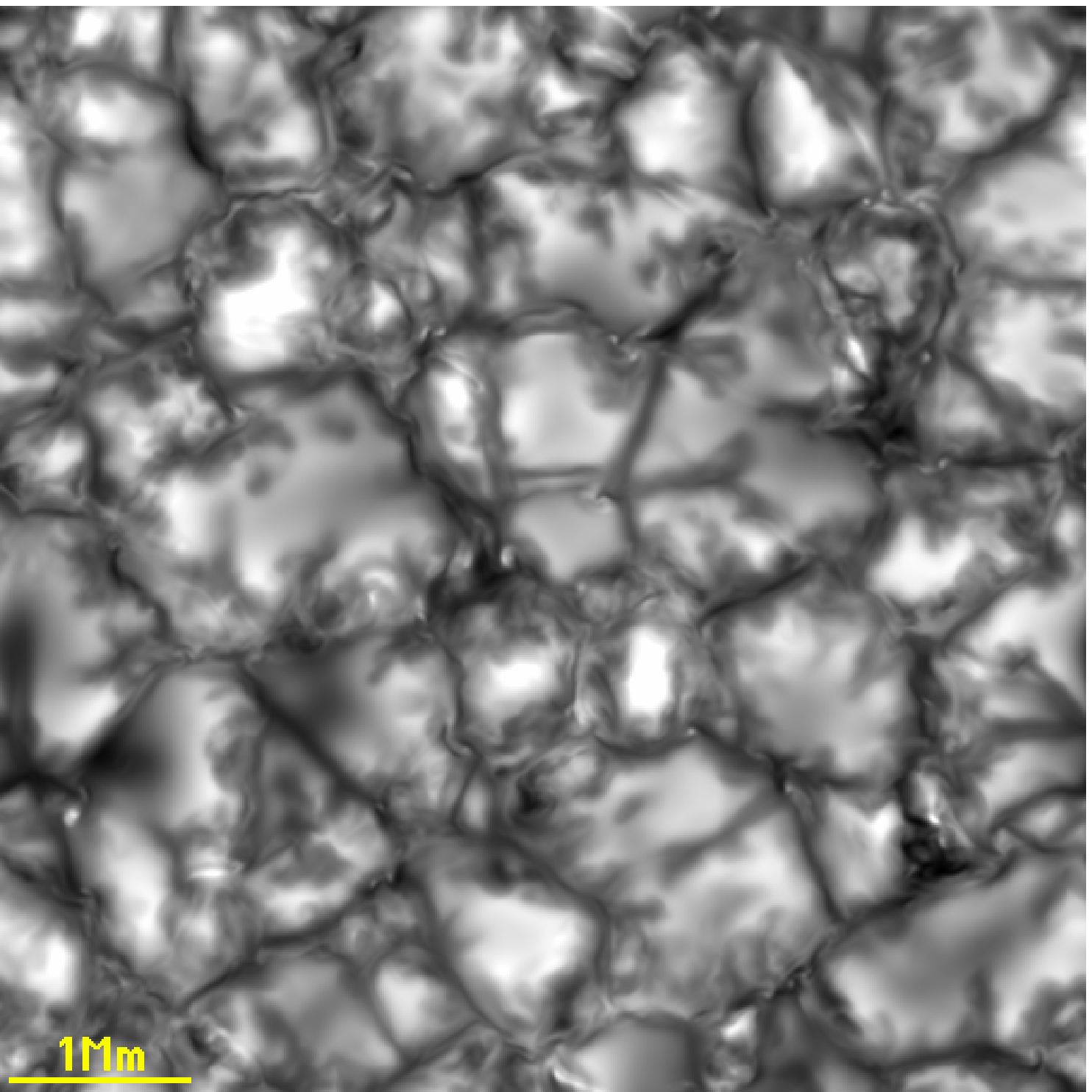} & \includegraphics[width=0.40\linewidth, trim=0mm -1mm 0mm 0mm]{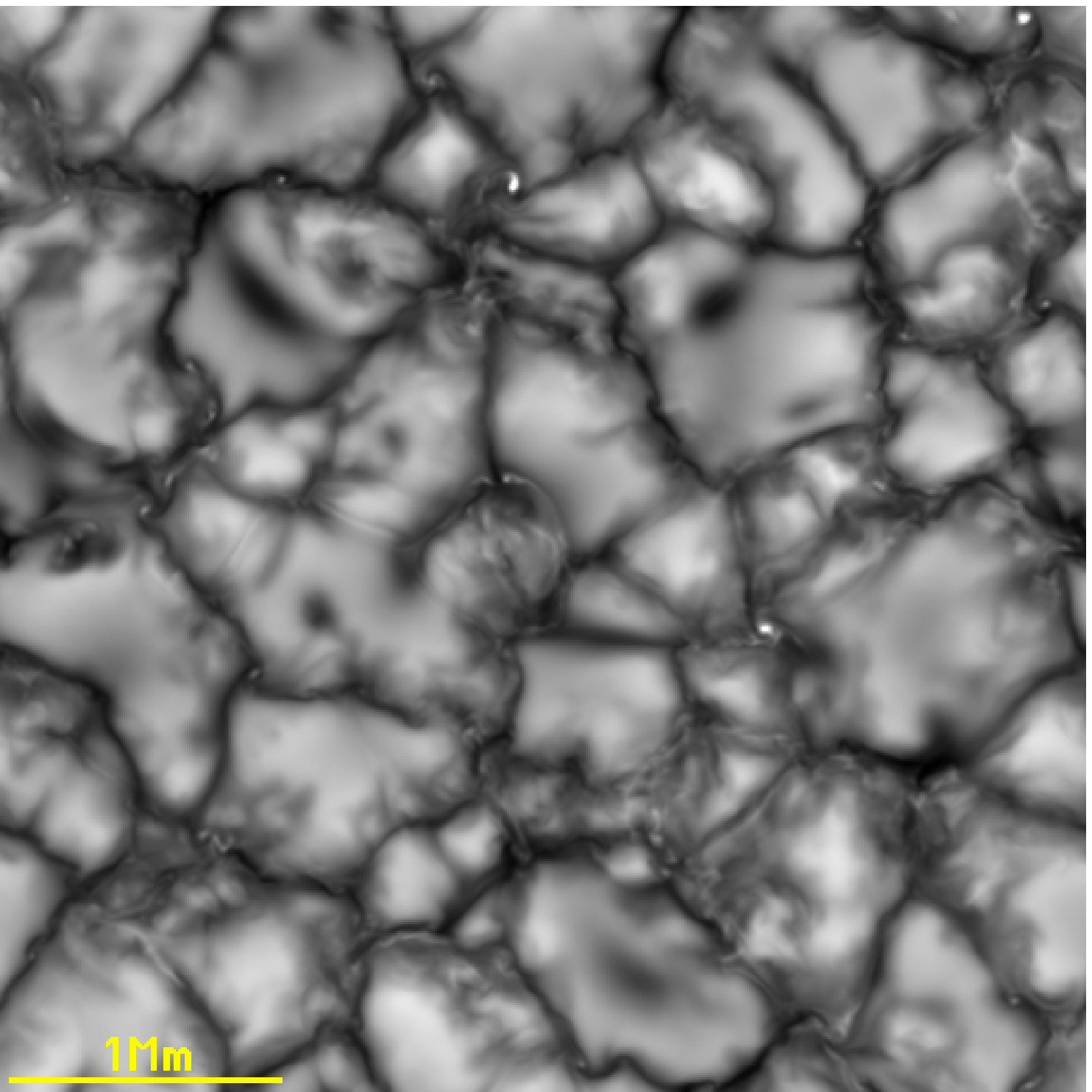}\\\hline

                                      {\bf\large M0V}  & {\bf\large M2V} \\ 
                                      \includegraphics[width=0.40\linewidth, trim=0mm -1mm 0mm 0mm]{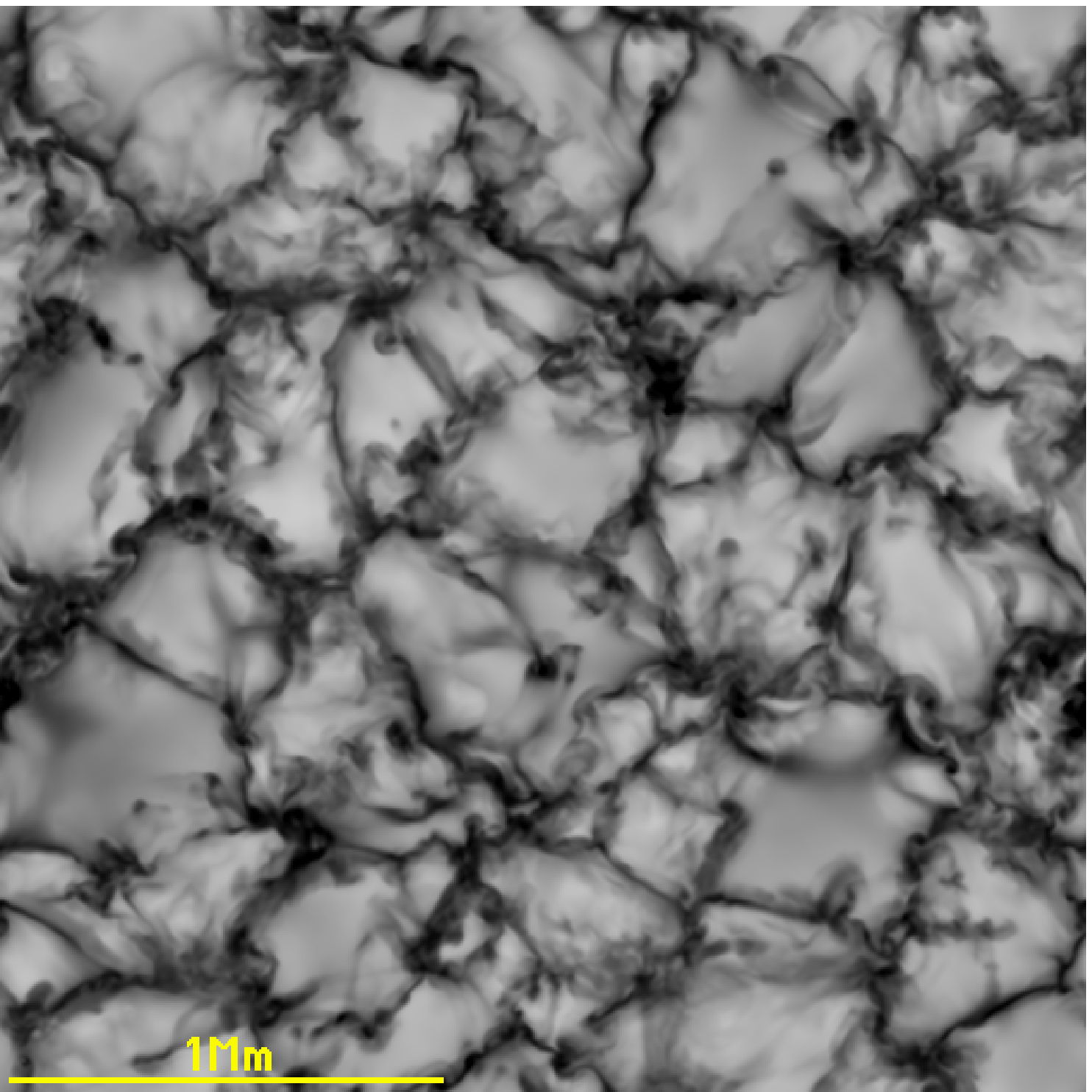}& \includegraphics[width=0.40\linewidth, trim=0mm -1mm 0mm 0mm]{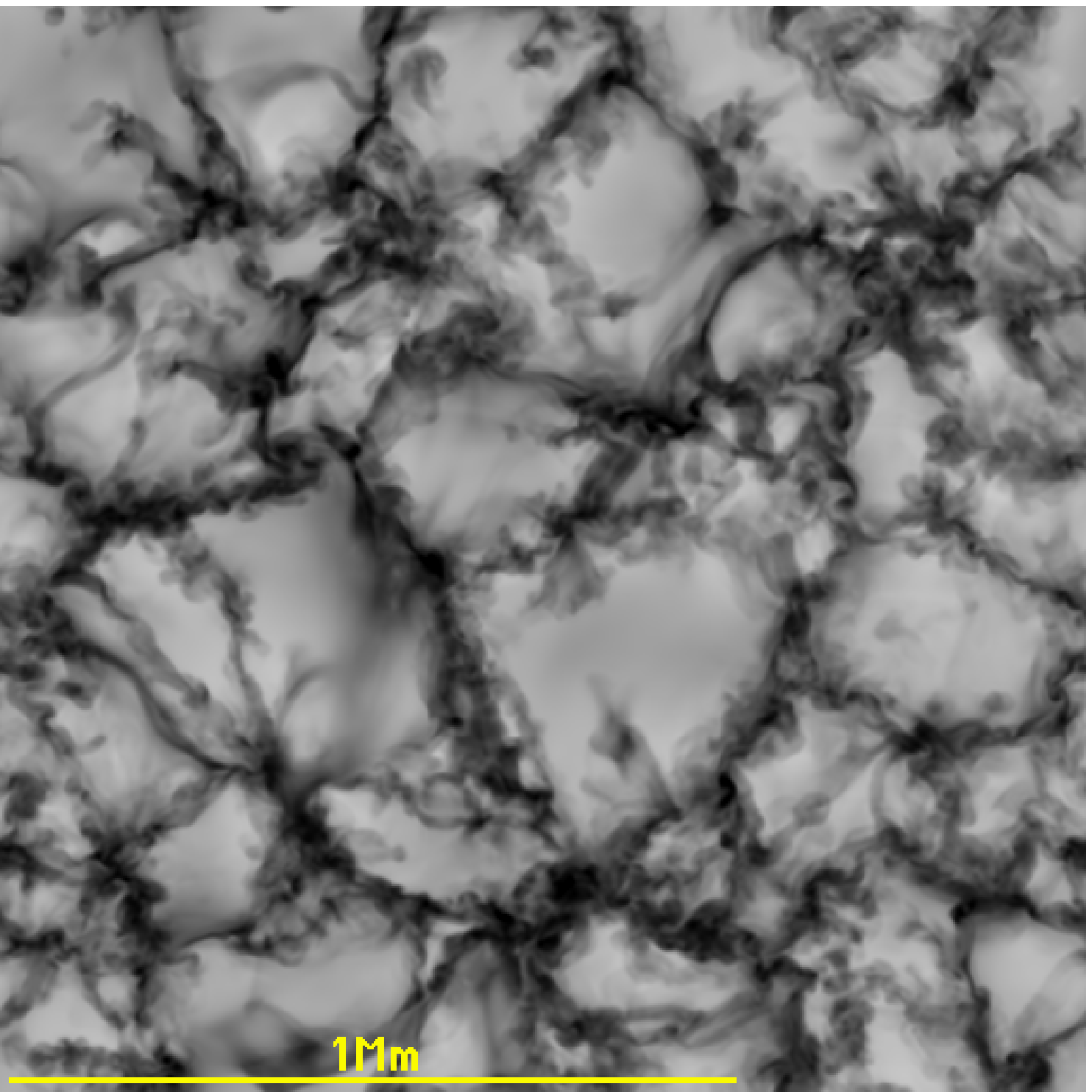}\\
    \end{tabular}
  \end{center}
  \caption{Images of the bolometric intensity computed from snapshots of six models with stellar parameters according to spectral types F3V, G2V, K0V, K5V, M0V, and M2V without magnetic field. Note the different box scale indicated by the yellow bar. The (RMS) intensity contrast (displayed enhanced) is 20.6\%, 15.4\%, 8.0\%, 7.1\%, 3.5\%, and 2.1\%, respectively.}\label{fig:convection}
\end{figure}
The MHD code MURaM is a ``box-in-the-star'' code that solves the equations of (non-ideal) MHD in three spatial dimensions with constant gravitational acceleration. It includes the relevant physical processes such as compressibility, partial ionization, and non-grey radiative energy transport. Hyperdiffusivities and artificial viscosities are introduced to account for energy dissipation on unresolved length scales. For more details, see \cite{MURaM1}.\\
To fit the surface conditions for different stellar spectral types, gravity and effective temperature were adjusted and the opacity bins were recalculated. The size of the computational box and the spatial resolution were modified in order to cover the relevant length scales of the different convection patterns.\\
The model grid comprises stars of spectral types F3V ({$T_{\mathrm{eff}}=6780\,\mathrm{K}$}, {$\log g[\mathrm{cgs}]=4.301$}), G2V (Sun;{$T_{\mathrm{eff}}=5770\,\mathrm{K}$}, {$\log g[\mathrm{cgs}]=4.438$}), K0 ({$T_{\mathrm{eff}}=4950\,\mathrm{K}$}, {$\log g[\mathrm{cgs}]=4.609$}), K5 ({$T_{\mathrm{eff}}=4370\,\mathrm{K}$}, {$\log g[\mathrm{cgs}]=4.699$}), M0 ({$T_{\mathrm{eff}}=3910\,\mathrm{K}$}, {$\log g[\mathrm{cgs}]=4.826$}), and M2 ({$T_{\mathrm{eff}}=3690\,\mathrm{K}$}, {$\log g[\mathrm{cgs}]=4.826$}) with solar chemical composition.\\
The start models were run with $\vec B\equiv \vec 0$ until a quasi-stationary state was reached. Then, a homogeneous vertical magnetic field with the field strength $B_0=20\,\mathrm{G}$, $100\,\mathrm{G}$, or $500\,\mathrm{G}$ was introduced.
\section{Results}
\subsection{General granulation characteristics}
\begin{figure}[!t]%
\begin{center}%
\begin{tabular}{cc}
\includegraphics[width=0.40\linewidth]{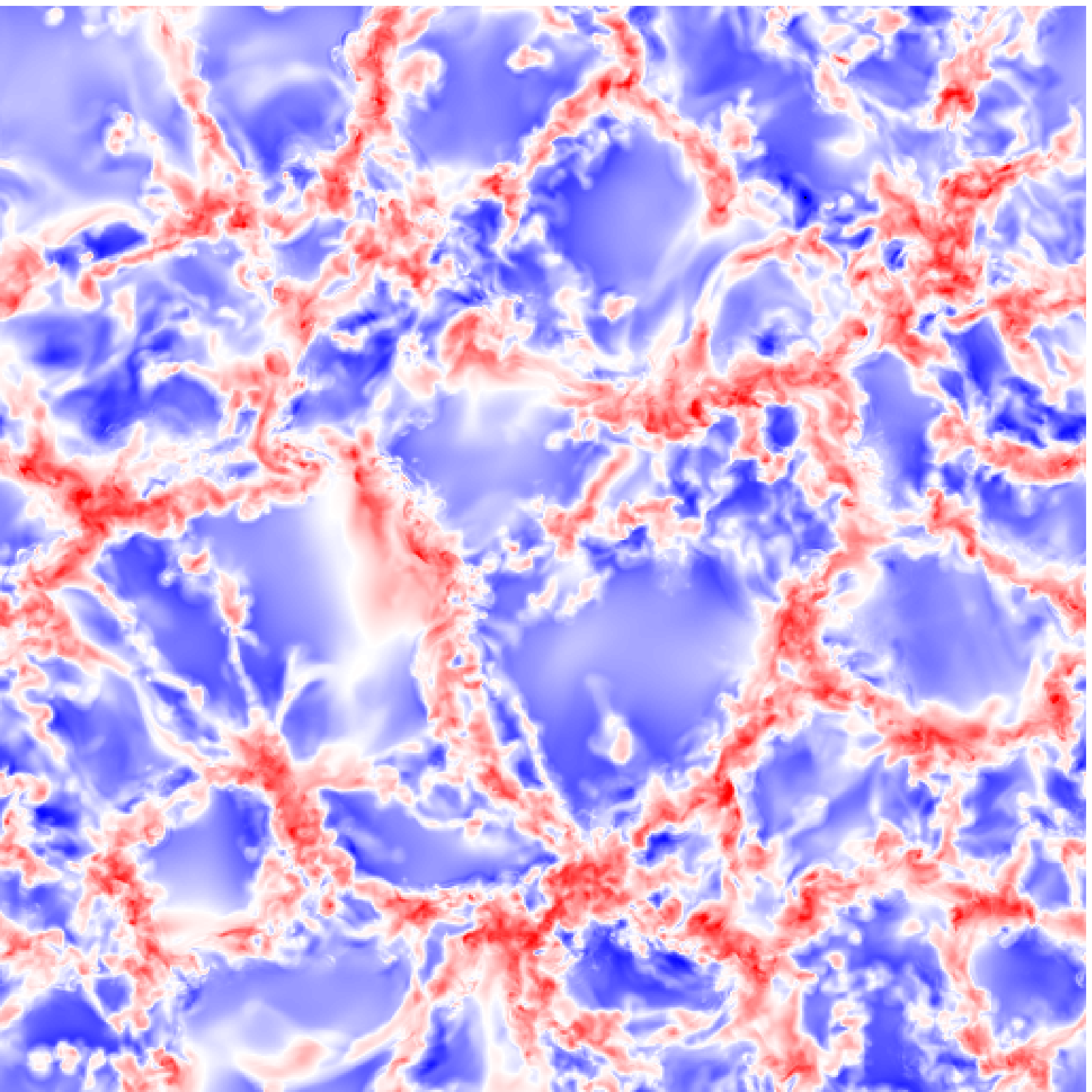} & \includegraphics[width=0.40\linewidth, trim=-15mm -10mm 0mm -5mm]{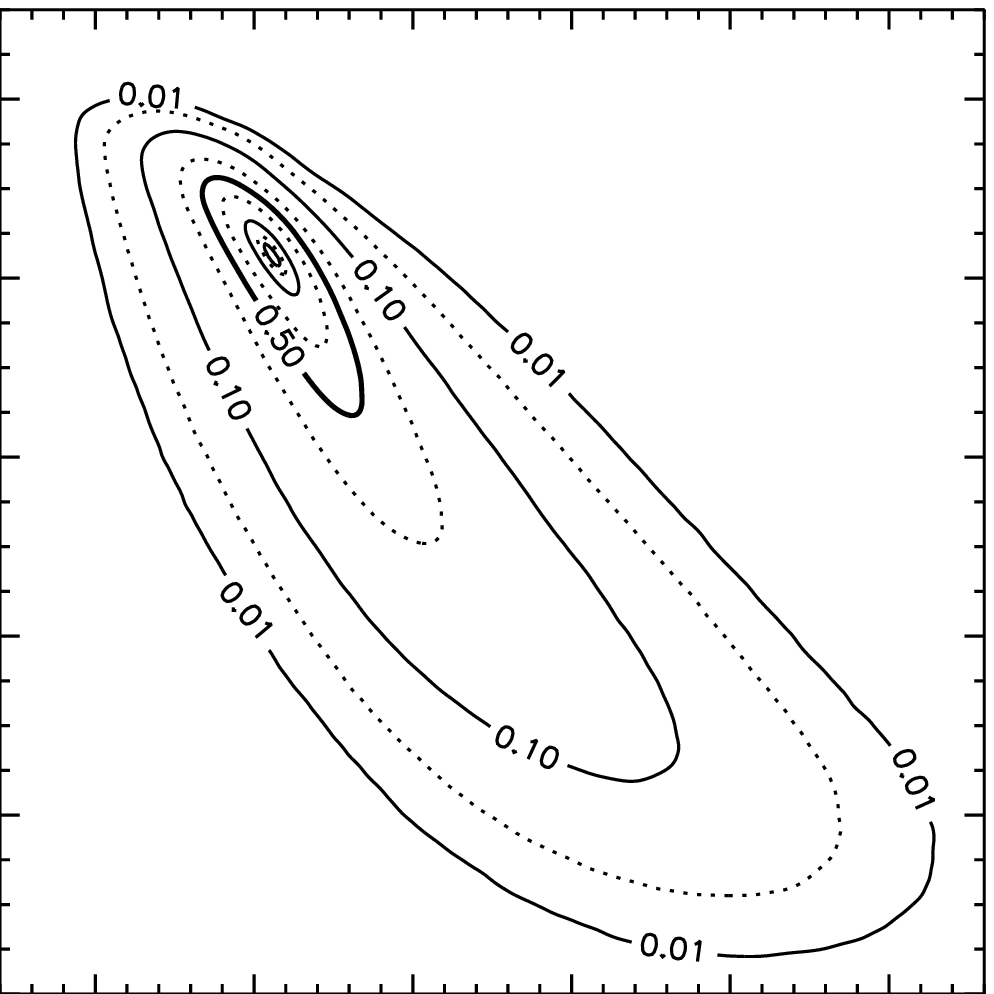}\\
\includegraphics[width=0.40\linewidth]{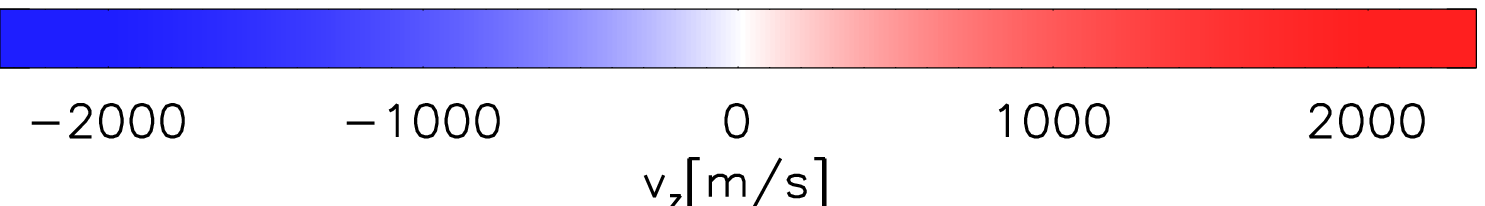} & ~ \\
\end{tabular}
\end{center}%

\caption{ Vertical velocities $v_z(\tau_{\mathrm{R}}\approx 1)$ at the optical surface computed from a snapshot of the M2V-star model (left panel; same time step as in Fig.\,\ref{fig:convection}). The right panel shows the distribution of bolometric intensity $I_{\mathrm{bol}}(\mu=1)$ as a function of $v_{z}(\tau_{\mathrm{R}}=1)$ in a contour plot: contour lines indicate isolines of point density in the scatter plot and are drawn for 0.01, 0.03, 0.1, 0.3, 0.5, 0.7, 0.9, 0.97, and 0.99 times the maximum point density.}\label{fig:velocities}
\end{figure}

Fig.\,\ref{fig:convection} shows intensity images of snapshots from the six models without magnetic field. The typical size of the granules scales roughly with the pressure scale height and thus decreases from more than 5\,Mm for the F3V model to less than 0.5\,Mm for M2V. The bolometric intensity contrast decreases from 20\% (F3V) to about 2\% (M2V).\\
As illustrated in Fig.\,\ref{fig:velocities}, the bright granules are upflows which carry hotter (brighter) gas to the optical surface, whereas the darker lanes between the granules are downflow regions where the radiatively cooled gas sinks back into the convective envelope. The typical vertical velocities decrease from about 7\,km\,s$^{-1}$ (F3V) to about 1\,km\,s$^{-1}$ (M2V). 

\subsection{Differences in the magneto-convection between G-type and M-type dwarfs} 

\begin{figure}[t]
  \begin{center}
    \begin{tabular}{cc}
      \multicolumn{2}{c}{{\normalsize $I_{\mathrm{bol}}(\mu=1)$} \quad\qquad {\large\bf G2V} \quad\qquad {\normalsize $B_z(\tau_{\mathrm{R}}\approx 1)$} }\\
      \includegraphics[width=0.42\linewidth, trim=0mm -2mm 0mm 0mm]{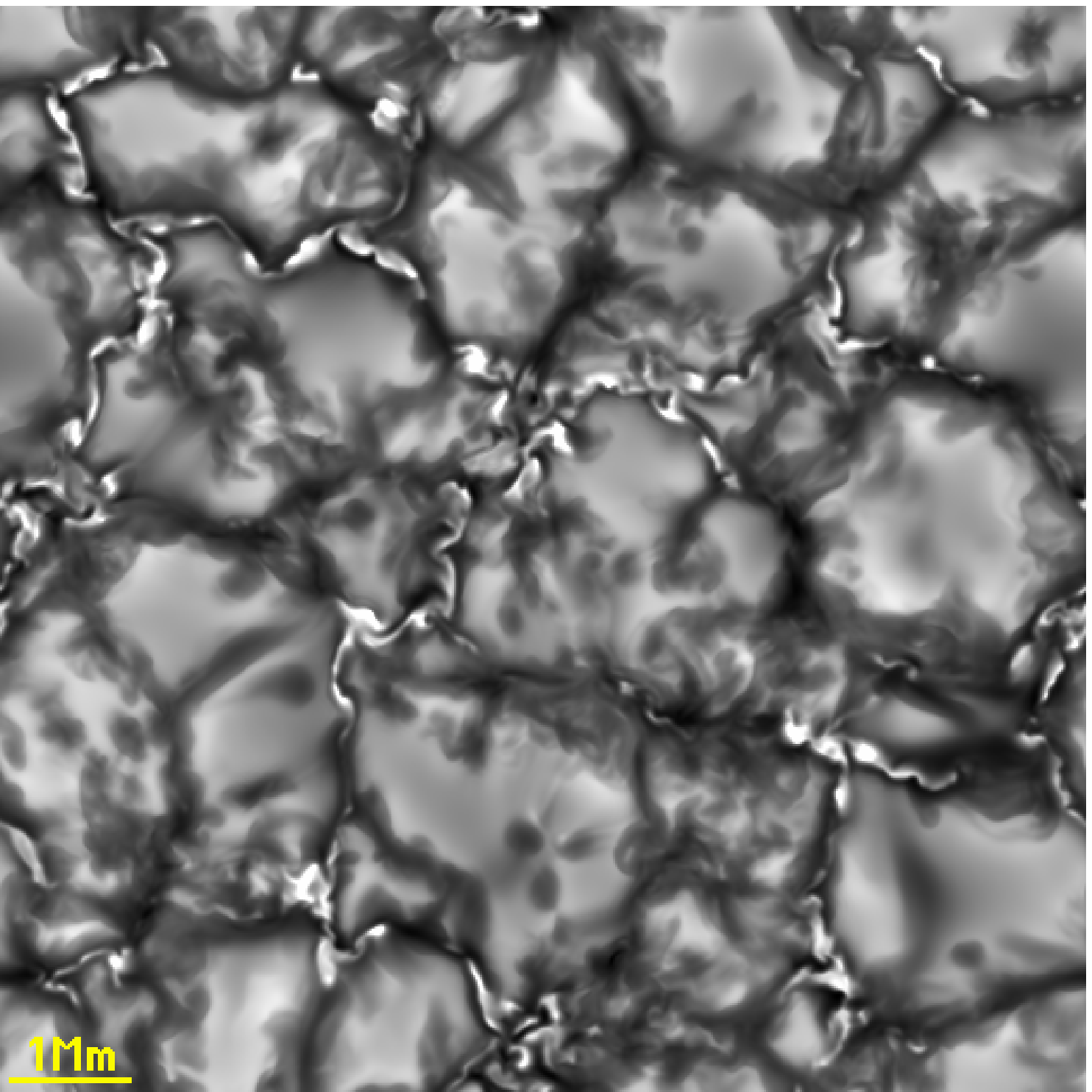}&\includegraphics[width=0.42\linewidth, trim=0mm -2mm 0mm 0mm]{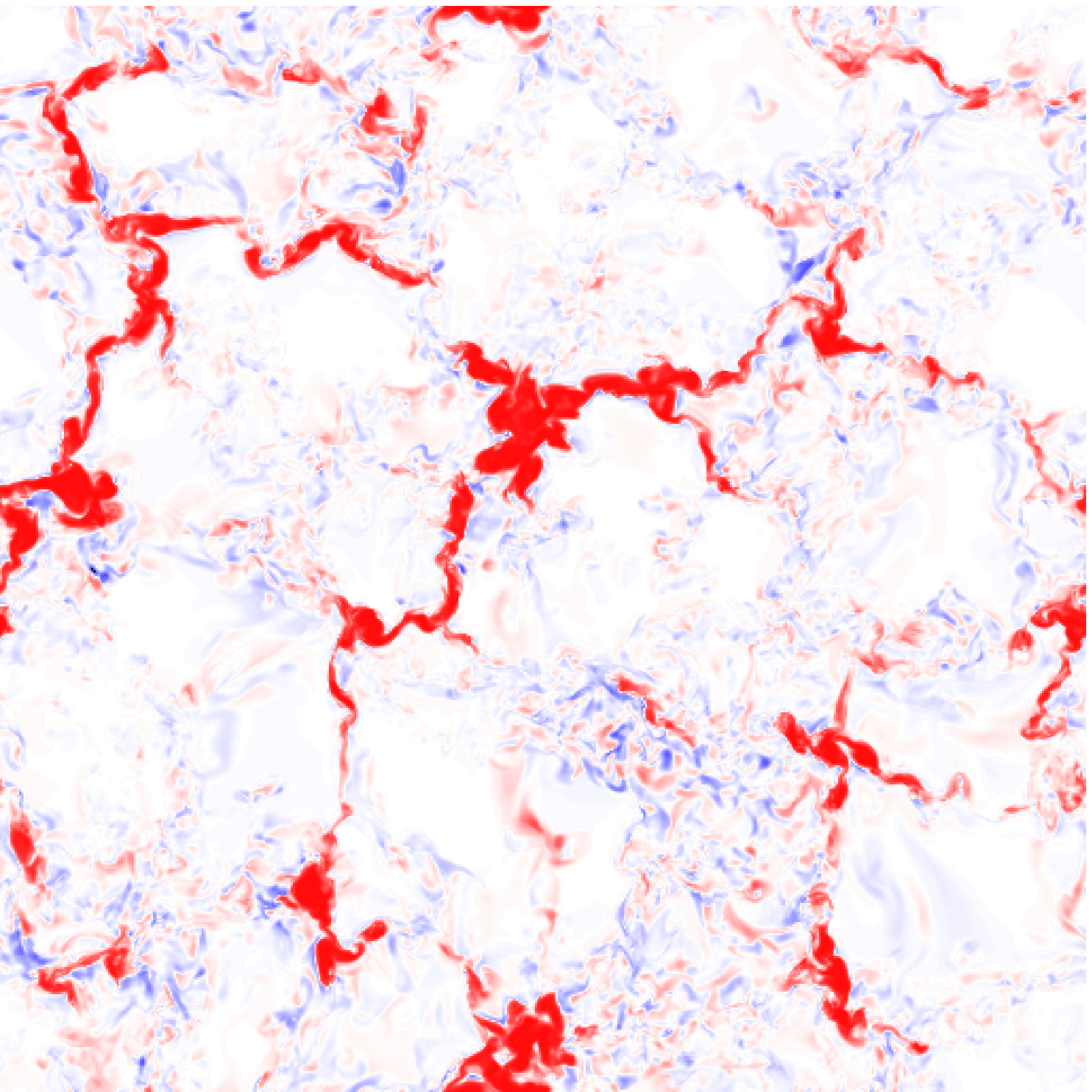}\\\hline
        \multicolumn{2}{c}{{\normalsize $I_{\mathrm{bol}}(\mu=1)$} \quad\qquad {\large\bf M2V} \quad\qquad {\normalsize $B_z(\tau_{\mathrm{R}}\approx 1)$}}\\
        \includegraphics[width=0.42\linewidth, trim=0mm -2mm 0mm 0mm]{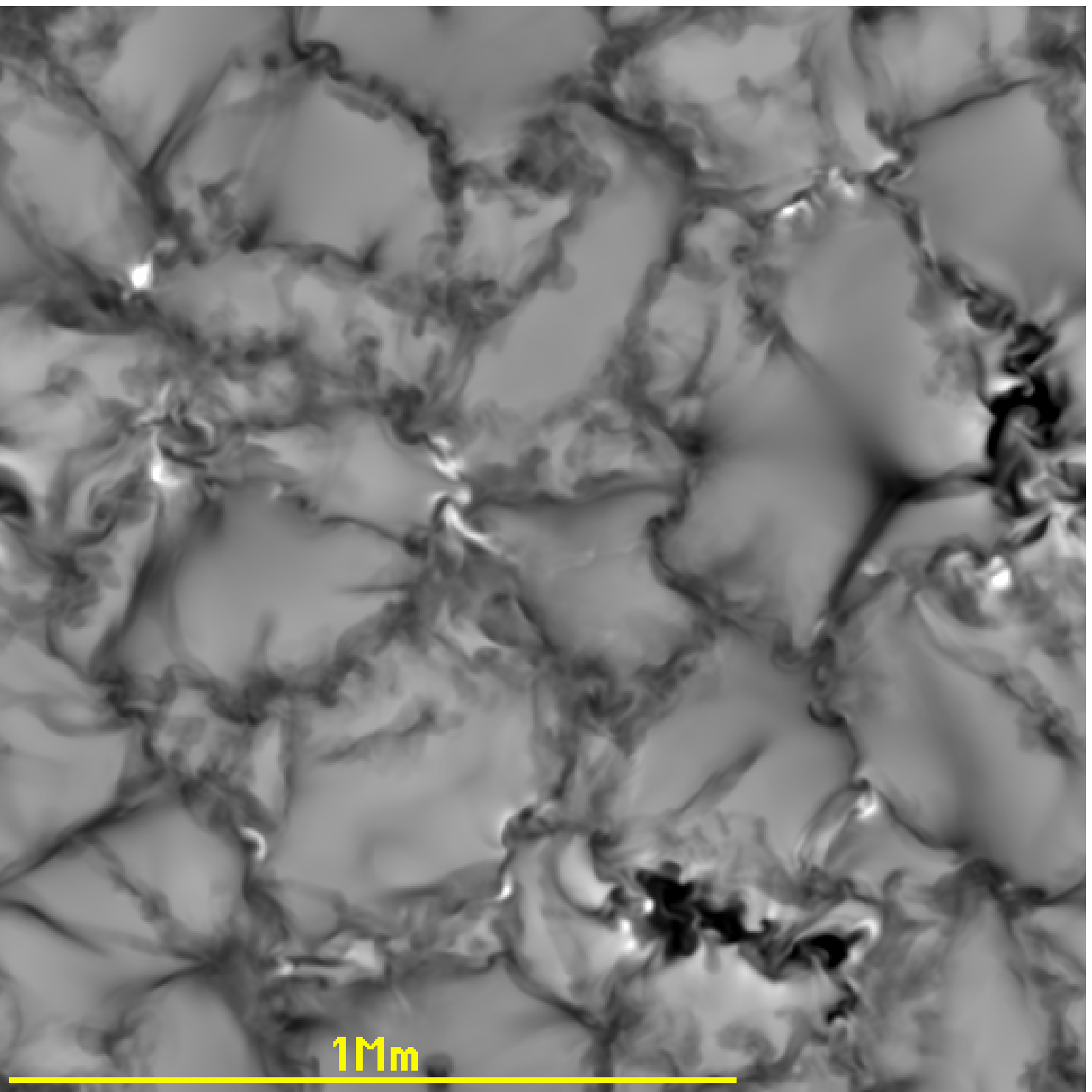}&\includegraphics[width=0.42\linewidth, trim=0mm -2mm 0mm 0mm]{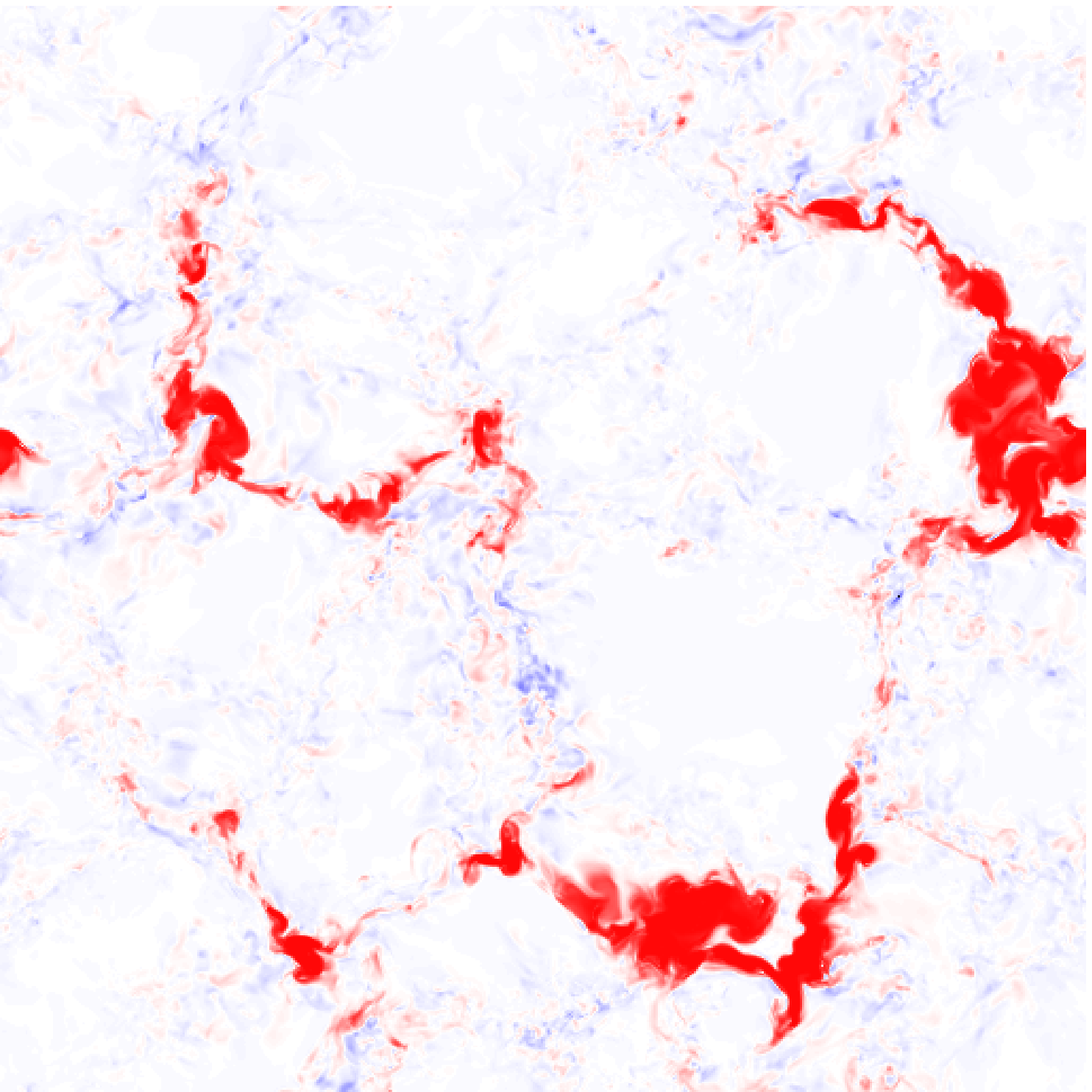}\\ 
        ~&\includegraphics[width=0.42\linewidth]{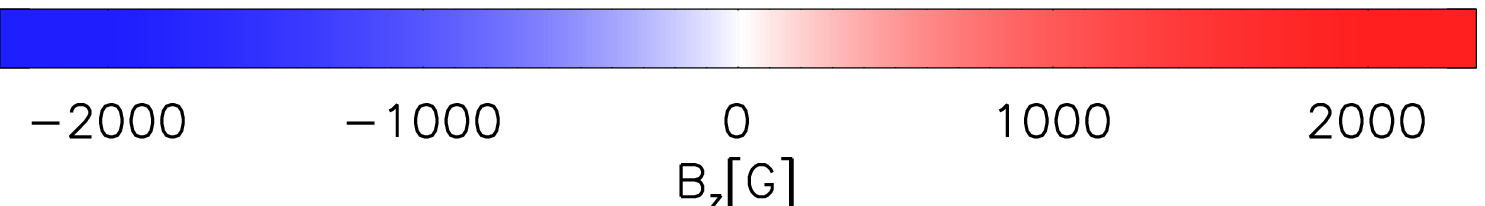}\\
      \end{tabular}
    \end{center}
    \caption{Snapshots of the bolometric intensity (left panels) and the vertical component of the magnetic field at the optical surface of the G2V and M2V star models. The initial field was $B_0=100\,\mathrm{G}$. The bolometric intensity contrast (displayed enhanced) is 14.9\% and 2.30\%.}\label{fig:mag.comp}
  \end{figure}

The modelled magneto-convection shows significant differences between
M-dwarfs and stars of earlier spectral types, e.\,g. the Sun. As
illustrated by Fig.~\ref{fig:mag.comp}, the initially homogeneous
magnetic flux is accumulated into very few structures of high field
strength, the cause of which are stable downflows.\\
\begin{figure}[t]
  \begin{center}
    \begin{tabular}{cc}
      \multicolumn{2}{c}{$B_0=20\,\mathrm{G}$}\\
      \includegraphics[width=0.399\linewidth]{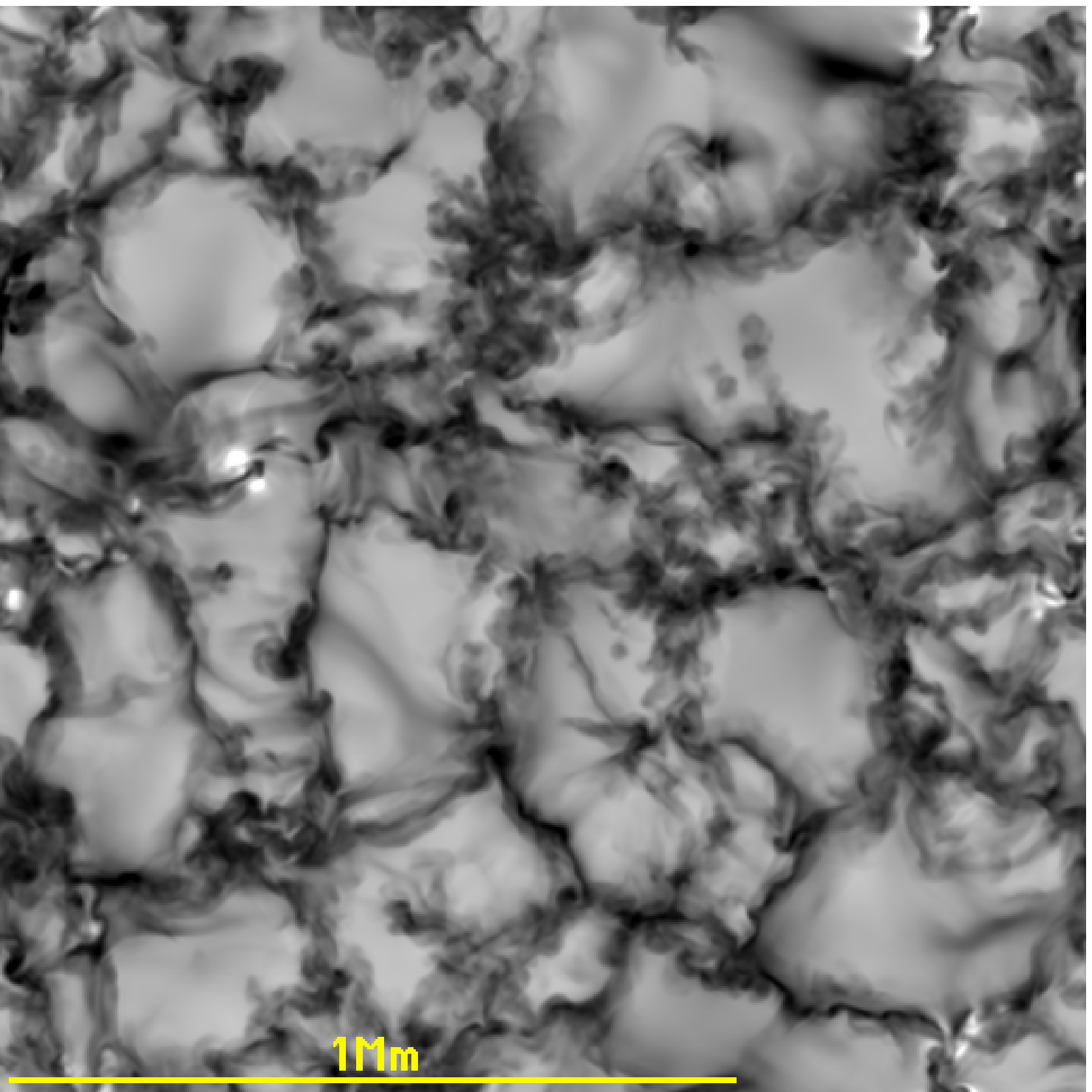} & \includegraphics[width=0.399\linewidth, trim=-15mm -10mm 0mm -5mm]{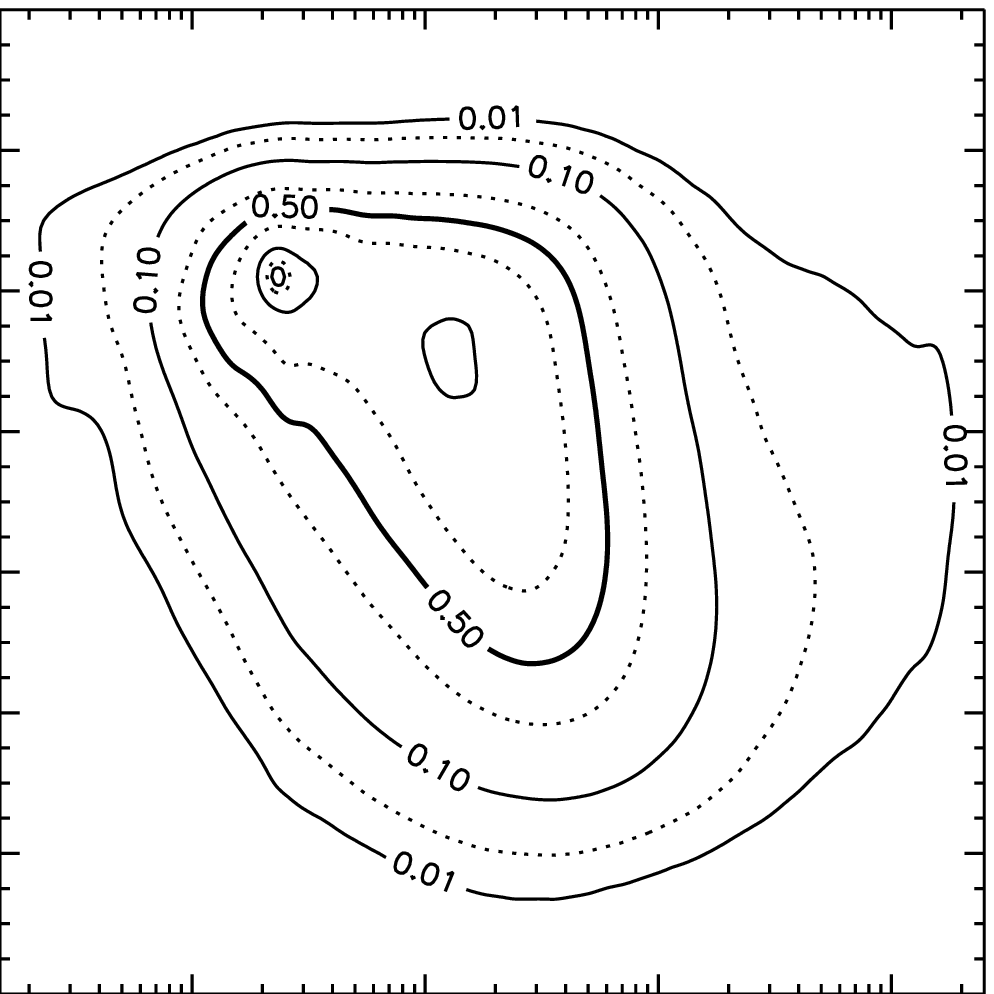}\\\hline
                      \multicolumn{2}{c}{$B_0=100\,\mathrm{G}$}\\
	              \includegraphics[width=0.399\linewidth]{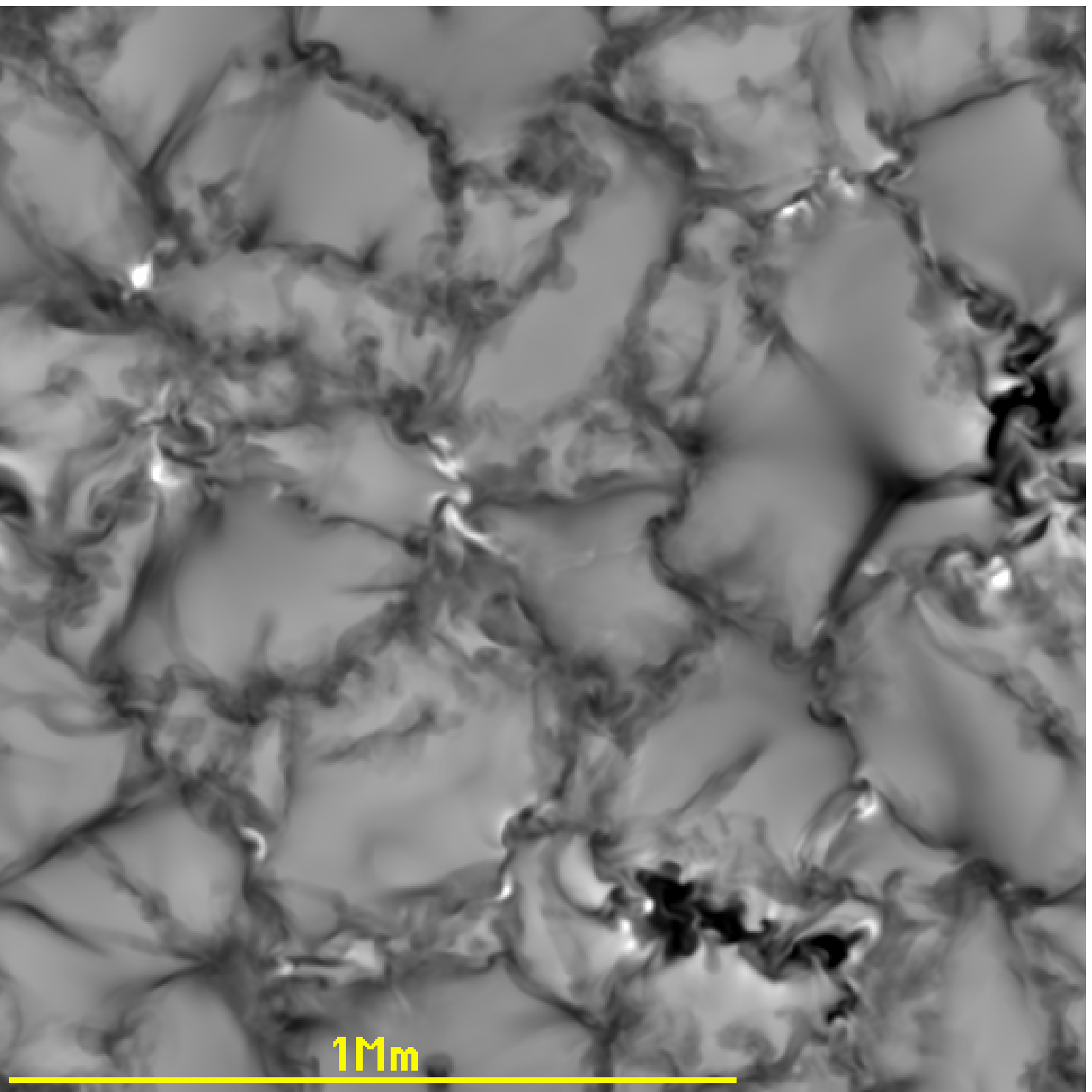} & \includegraphics[width=0.399\linewidth, trim=-15mm -10mm 0mm -5mm]{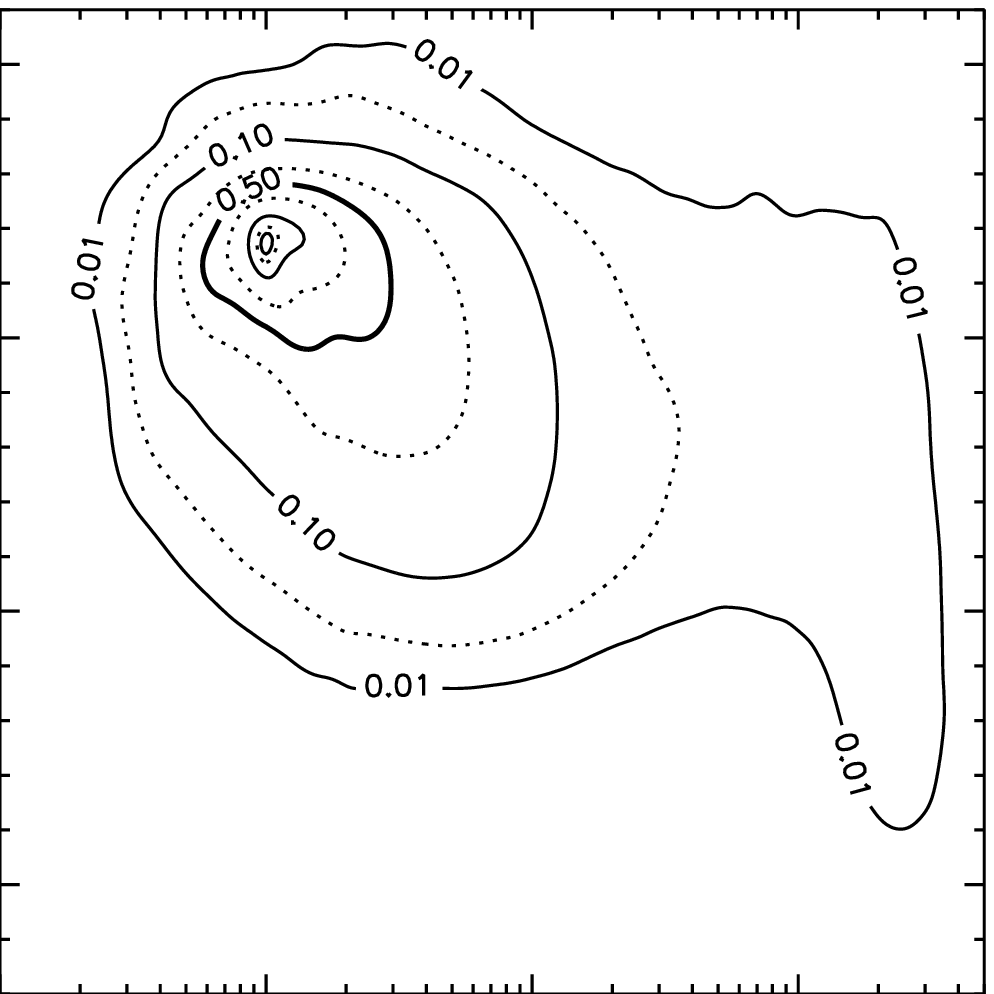}\\\hline
                                      \multicolumn{2}{c}{$B_0=500\,\mathrm{G}$}\\
	                              \includegraphics[width=0.399\linewidth]{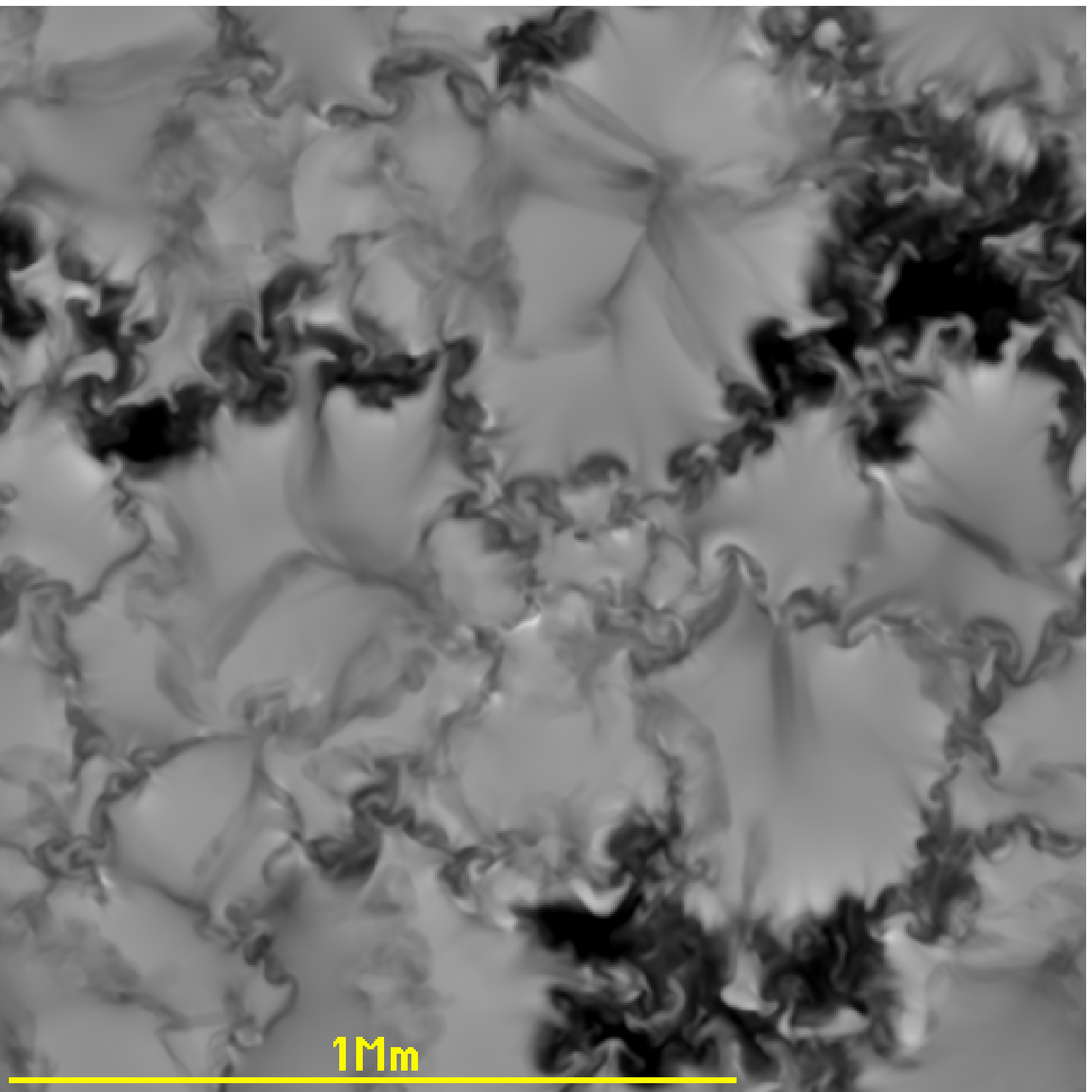} & \includegraphics[width=0.399\linewidth, trim=-15mm -10mm 0mm -5mm]{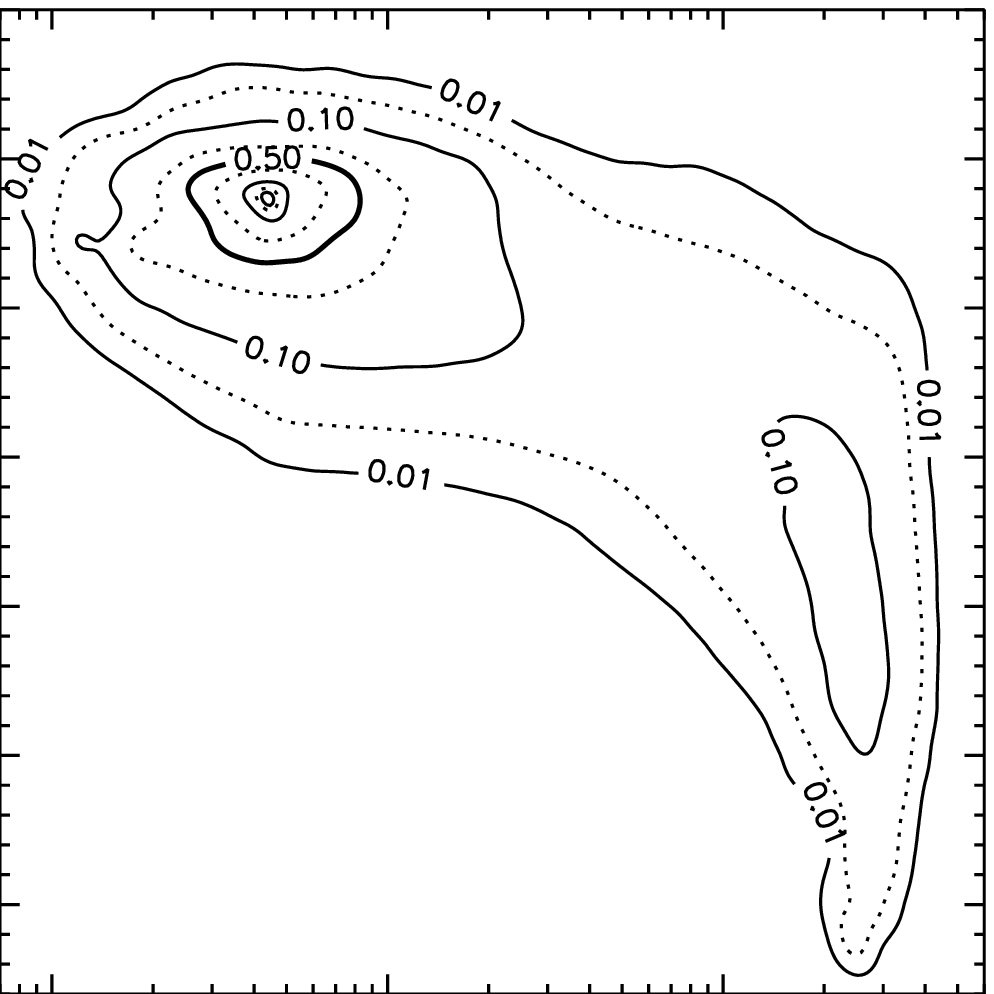}\\
    \end{tabular}
  \end{center}
  \caption{Snapshots of the bolometric intensity of the M2V
    model with different initial fields $B_0=20,100,500\,\mathrm{G}$
    (left panels). The intensity contrast (displayed enhanced) is
    2.11\%, 2.37\%, and 5.26\% (from top to bottom). The right
    panels show the according distribution of bolometric
    intensity $I_{\mathrm{bol}}(\mu=1)/\langle I_{\mathrm{bol}}(\mu=1)
    \rangle$ as a function of field strength $|B(\tau\approx
    1)|$. Contour lines indicate isolines of the point densities in
    the scatter plot and are drawn for 0.01, 0.03, 0.1, 0.3, 0.5, 0.7, 0.9,
    0.97, and 0.99 times the maximum point
    density.\label{fig:mag.20100500}}
\end{figure}%
While solar magnetic structures appear as bright features,
the magnetic structures on M-dwarfs tend to be rather dark. In the
case of the Sun, magnetic structures create a strong depression of the
optical surface with hot side walls that can radiatively heat the
interior of the magnetic structure. This mechanism is much less
efficient for the magnetic structures in M-dwarf atmospheres: owing to
the much higher densities, the depressions formed by the magnetic
structures are very shallow and their side walls have small excess
temperatures due to a shallow temperature gradient. Since the magnetic field suppresses
convective energy transport, the structures cool down. As
Fig.~\ref{fig:mag.20100500} shows, magnetic structures in our M2V-star
model tend to be darker than the environment at all initial fields
$B_0$, the effect being much more pronounced for strong plage fields
($B_0=500\,\mathrm{G}$).\\ This indicates that plage regions on M-stars
might not show bright points but rather ``pores'' and small ``star
spots'' of reduced intensity, which has a crucial impact on the
interpretation of observational data such as M-dwarf spectra.\\

\acknowledgements We would like to thank R. Cameron for help with setting up the MURaM simulations.

\bibliography{beeck_b}

\begin{thebibliography}{}
\expandafter\ifx\csname natexlab\endcsname\relax\def\natexlab#1{#1}\fi
\expandafter\ifx\csname url\endcsname\relax
  \def\url#1{\texttt{#1}}\fi
\expandafter\ifx\csname urlprefix\endcsname\relax\def\urlprefix{URL }\fi
\providecommand{\eprint}[2][]{\url{#2}}

\bibitem[{{V{\"o}gler} et~al.(2005){V{\"o}gler}, {Shelyag}, {Sch{\"u}ssler},
  {Cattaneo}, {Emonet}, \& {Linde}}]{MURaM1}
{V{\"o}gler}, A., {Shelyag}, S., {Sch{\"u}ssler}, M., {Cattaneo}, F., {Emonet},
  T., \& {Linde}, T. 2005, \aap, 429, 335

\end{thebibliography}

\end{document}